\documentstyle[aps,prl,epsfig]{revtex} 

\def \kl           {\mbox{$K_L$}}

\def \kp           {\mbox{$K^+$}}
\def \klpill       {\mbox{$K_L \rightarrow \pi^0 l \bar{l}$}}
\def \klpiee       {\mbox{$K_L \rightarrow \pi^0 e^+ e^-$}}
\def \klpimumu     {\mbox{$K_L \rightarrow \pi^0 \mu^+ \mu^-$}}
\def \klpinunu     {\mbox{$K_L \rightarrow \pi^0 \nu \bar{\nu}$}}

\def \kleegg       {\mbox{$K_L \rightarrow e^+ e^- \gamma \gamma$}}
\def \klmumug      {\mbox{$K_L \rightarrow \mu^+ \mu^- \gamma$}}
\def \klmumugg     {\mbox{$K_L \rightarrow \mu^+ \mu^- \gamma \gamma$}}
\def \klelelgg     {\mbox{$K_L \rightarrow l^+ l^- \gamma \gamma$}}

\def \klpmz        {\mbox{$K_L \rightarrow \pi^+ \pi^- \pi^0$}}

\def \klethree     {\mbox{$K_L \rightarrow \pi^\pm e^\mp \nu$}}
\def \klmthree     {\mbox{$K_L \rightarrow \pi^\pm \mu^\mp \nu$}}

\def \klpm         {\mbox{$K_L \rightarrow \pi^+ \pi^-$}}

\def \piz          {\mbox{$\pi^0$}}

\def \pipm         {\mbox{$\pi^\pm$}}

\def \bra#1        {\mbox{$\left\langle #1 \right|$}}
\def \ket#1        {\mbox{$\left| #1\right\rangle$}}
\def \VEV#1        {\mbox{$\left\langle #1\right\rangle$}}
\def \braket#1#2   {\mbox{$\left\langle #1 \left| #2\right\rangle$}}

\def \itapt        {\mbox{$P_\perp$}}
\def \itaptsq      {\mbox{$P_\perp^2$}}
\def \BR#1         {\mbox{${\mathcal B}$(#1)}}  

\def \partder#1#2  {\mbox{$\partial #1\over\partial #2$}}
\def \secder#1#2#3 {\mbox{$\partial^2 #1\over\partial #2 \partial #3$}}
\def \um           {\mbox{\,$\mu$m}}
\def \cm           {\mbox{\,cm}}
\def \mm           {\mbox{\,mm}}

\def \mev          {\mbox{\,MeV}}
\def \mevc         {\mbox{\,MeV/c}}
\def \mevcc        {\mbox{\,MeV/c$^2$}}
\def \gev          {\mbox{\,GeV}}
\def \gevc         {\mbox{\,GeV/c}}

\def \mrad         {\mbox{\,mrad}}

\def \vs           {\mbox{\it vs.}}
\def \eg           {\mbox{\it e.g.}}     

\def \etal         {\mbox{\it et al.}}

\def \EPJ          {\mbox{European Phys. Journal}}

\def \NIM          {\mbox{Nucl. Instr. Meth.}}
\def \NP           {\mbox{Nucl. Phys.}}

\def \PR           {\mbox{Phys. Rev.}}
\def \PRL          {\mbox{Phys. Rev. Lett.}}

\def \RMP          {\mbox{Rev. Mod. Phys.}}

\begin{document}
\draft          
\title{Search for the Decay $K_L \rightarrow \pi^0 \mu^+ \mu^-$}
\date{\today}
\maketitle

\begin{center}\normalsize\parindent=0.in
A.~Alavi-Harati$^{12}$,
I.F.~Albuquerque$^{10}$,
T.~Alexopoulos$^{12}$,
M.~Arenton$^{11}$,
K.~Arisaka$^2$,
S.~Averitte$^{10}$,
A.R.~Barker$^5$,
L.~Bellantoni$^{7,\dagger}$,
A.~Bellavance$^9$,
J.~Belz$^{10}$,
R.~Ben-David$^7$,
D.R.~Bergman$^{10}$,
E.~Blucher$^4$, 
G.J.~Bock$^7$,
C.~Bown$^4$, 
S.~Bright$^4$,
E.~Cheu$^1$,
S.~Childress$^7$,
R.~Coleman$^7$,
M.D.~Corcoran$^9$,
G.~Corti$^{11}$, 
B.~Cox$^{11}$,
M.B.~Crisler$^7$,
A.R.~Erwin$^{12}$,
R.~Ford$^7$,
A.~Glazov$^4$,
A.~Golossanov$^{11}$,
G.~Graham$^4$, 
J.~Graham$^4$,
K.~Hagan$^{11}$,
E.~Halkiadakis$^{10}$,
K.~Hanagaki$^8$,  
M.~Hazumi$^8$,
S.~Hidaka$^8$,
Y.B.~Hsiung$^7$,
V.~Jejer$^{11}$,
J.~Jennings$^2$,
D.A.~Jensen$^7$,
R.~Kessler$^4$,
H.G.E.~Kobrak$^{3}$,
J.~LaDue$^5$,
A.~Lath$^{10}$,
A.~Ledovskoy$^{11}$,
P.L.~McBride$^7$,
A.P.~McManus$^{11}$,
P.~Mikelsons$^5$,
E.~Monnier$^{4,*}$,
T.~Nakaya$^7$,
K.S.~Nelson$^{11}$,
H.~Nguyen$^7$,
V.~O'Dell$^7$, 
M.~Pang$^7$, 
R.~Pordes$^7$,
V.~Prasad$^4$, 
C.~Qiao$^4$, 
B.~Quinn$^4$,
E.J.~Ramberg$^7$, 
R.E.~Ray$^7$,
A.~Roodman$^4$, 
M.~Sadamoto$^8$, 
S.~Schnetzer$^{10}$,
K.~Senyo$^8$, 
P.~Shanahan$^7$,
P.S.~Shawhan$^4$,
W.~Slater$^2$,
N.~Solomey$^4$,
S.V.~Somalwar$^{10}$, 
R.L.~Stone$^{10}$, 
I.~Suzuki$^8$,
E.C.~Swallow$^{4,6}$,
R.A.~Swanson$^{3}$,
S.A.~Taegar$^1$,
R.J.~Tesarek$^{10}$, 
G.B.~Thomson$^{10}$,
P.A.~Toale$^5$,
A.~Tripathi$^2$,
R.~Tschirhart$^7$, 
Y.W.~Wah$^4$,
J.~Wang$^1$,
H.B.~White$^7$, 
J.~Whitmore$^7$,
B.~Winstein$^4$, 
R.~Winston$^4$, 
T.~Yamanaka$^8$,
E.D.~Zimmerman$^4$
\vspace*{0.1in}

$^1$ University of Arizona, Tucson, AZ 85721 \\
$^2$ University of California at Los Angeles, Los Angeles, CA 90095 \\
$^{3}$ University of California at San Diego, La Jolla, CA 92093 \\
$^4$ The Enrico Fermi Institute, The University of Chicago, 
Chicago, IL 60637 \\
$^5$ University of Colorado, Boulder, CO 80309 \\
$^6$ Elmhurst College, Elmhurst, IL 60126 \\
$^7$ Fermi National Accelerator Laboratory, Batavia, IL 60510 \\
$^8$ Osaka University, Toyonaka, Osaka 560 Japan \\
$^9$ Rice University, Houston, TX 77005 \\
$^{10}$ Rutgers University, Piscataway, NJ 08855 \\
$^{11}$ The Department of Physics and Institute of Nuclear and 
Particle Physics, University of Virginia, 
Charlottesville, VA 22901 \\
$^{12}$ University of Wisconsin, Madison, WI 53706 \\
$^{*}$ On leave from C.P.P. Marseille/C.N.R.S., France \\
$^{\dagger}$ To whom correspondence should be addressed.
\end{center}

\begin{abstract}
We report on a search for the decay \klpimumu\ carried out as a part
of the KTeV experiment at Fermilab.  This decay is expected to have a
significant $CP$\, violating contribution and a direct measurement will
either support the CKM mechanism for $CP$\, violation or point to new
physics.  Two events were observed in the 1997 data
with an expected background of $0.87 \pm 0.15$\ events, and we
set an upper limit
\BR{\klpimumu} $\,<3.8 \times 10^{-10}$\ at the 90\% confidence
level.
\end{abstract}

\pacs{13.20.Eb, 11.30.Er, 14.40.A}
%
%
%
%
The decays \klpill\ are interesting decays for the study of 
$CP$\, violation and can be used to search for new physics.
There are three expected contributions to the amplitude: a 
$CP$\, conserving contribution which proceeds through the 
$\pi^0 \gamma^* \gamma^*$\ intermediate state, an indirectly
$CP$\, violating contribution from
\mbox{$K_1 \rightarrow \pi^0 l \bar{l}$}, and a directly
$CP$\, violating contribution from electroweak penguin and $W$\
box diagrams \cite{ref:reviews,ref:Buras,ref:Donny}.  Branching ratio
predictions in theories containing exotic (\eg, SUSY) particles that 
contribute to the penguin amplitudes are significantly
higher \cite{ref:hope}.

The sizes of the three contributions depend on the flavor of the final
state lepton.  The greatest theoretical interest is in the
\klpinunu\ case, where the direct $CP$\, violating amplitude dominates
and a theoretically clean measurement \cite{ref:LL} of the Wolfenstein
\cite{ref:LW} parameter $\eta$\ should be possible.  However, measuring
a final state with two neutrinos and a single pion is experimentally
challenging, and the current experimental limit \cite{ref:Kazu} remains
four orders of magnitude above the Standard Model expectation of 
$\sim 3 \times 10^{-11}$.

In contrast, \klpiee\ and \klpimumu\ are comparatively straightforward
to detect, although all three amplitudes are present in these modes.  
This letter presents a new limit on \BR{\klpimumu\,} ; the existing
limit \cite{ref:Debbie} is $5.1 \times 10^{-9}$\, at the 90\% C.L.
For the data taken by KTeV in 1997 and discussed here, a single event 
observed in the muon mode would correspond to a branching ratio of
$7 \times 10^{-11}$, which approaches the Standard Model expectation
\cite{ref:Sehgal,ref:EPR} of 
\BR{\klpimumu} $\,\sim (0.44 - 1.00) \times 10^{-11}$.

\begin{figure}
\caption{The KTeV detector configuration for rare decay studies.  The
TRD, Hadron Veto, and the Photon Veto in the beamline were not used in
this analysis.}
\label{fig:detector}
\vspace{0.5in}
\epsfig{file=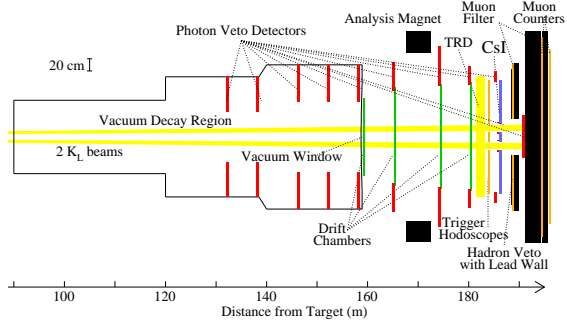,height=3.0in,angle=270}
\end{figure}

Figure \ref{fig:detector} shows a plan view of the KTeV detector, which
has been described elsewhere \cite{ref:Kazu,ref:det}.  An
800\gev\ proton beam, with typically $3.5 \times 10^{12}$\ protons
per 19\,s Fermilab Tevatron spill every minute, was targeted at a
vertical angle of 4.8\mrad\ on a 1.1 interaction length (30\cm) BeO
target. Photons were converted by 76\mm\ of lead immediately downstream
of the target.  Charged particles were then removed with magnetic
sweeping.  Collimators defined two 0.25\,$\mu$sr beams that entered
the KTeV apparatus 94\,m downstream of the target. The 65\,m vacuum
\mbox{($\sim10^{-6}$\,Torr)} decay region extended to the first drift
chamber.  The spectrometer consisted of a dipole magnet surrounded by
four (\mbox{$1.28 \times 1.28$\,m$^2$} to 
\mbox{$1.77 \times 1.77$\,m$^2$}) drift chambers with $\sim$100\um\
position resolution in both horizontal and vertical directions.
Helium filled bags occupied the spaces between the drift chambers; the
magnetic field imparted a $\pm$205\mevc\ horizontal momentum kick.  The
spectrometer had a momentum resolution of
$\sigma(P)/P = 0.38\% \oplus 0.016\% P$, where $P$\ is in\,\gevc.  The
electromagnetic calorimeter consisted of 3100 pure CsI crystals.
Each crystal was 50\cm\ (27 radiation lengths, 1.4 interaction lengths)
long.  Crystals in the central \mbox{$1.2 \times 1.2$\,m$^2$} section
of the calorimeter had a cross-sectional area of
\mbox{$2.5 \times 2.5$\cm$^2$}, and those in the outer region (out
to \mbox{$1.9 \times 1.9$\,m$^2$}) had a \mbox{$5 \times 5$\cm$^2$}
area.  The calorimeter's energy resolution for photons was
$\sigma(E)/E = 0.45\% \oplus 2\%/\sqrt{E}$, where $E$\ is in\.\gev,
and its position resolution was $\sim$1\mm.  Nine photon veto
assemblies (lead scintillator sandwiches) detected particles leaving
the fiducial volume.  Two scintillator hodoscopes in front of the
calorimeter were used to trigger on charged particles.  The hodoscopes
and the calorimeter had two holes (\mbox{$15 \times 15$\cm} at the
calorimeter) to let the neutral beams pass through without interaction.
The muon filter, located behind the calorimeter, was constructed of a
10\cm\ thick lead wall followed by three steel walls totaling
511\cm\ thickness.  Scintillator planes with 15\cm\ segmentation in
both horizontal and vertical directions (MU3) were located after the
third steel wall.  The segmentation was comparable to the multiple
scattering angle of 10\gev\ muons at MU3.  Pion punch-through
probabilities, including decays downstream of the calorimeter, 
were taken as a function of momentum from \klethree\, data and are
on the order of $2 \times 10^{-3}$.  The data acquisition system
reconstructed events online, and the results were used to filter
the data.

The signature we searched for is two tracks from oppositely charged
particles with a common vertex that deposit little energy in the
calorimeter and created two hits at MU3 from these muons.  The
\piz\ creates two electromagnetic showers in the calorimeter with
$m_{\gamma\gamma} = m_{\pi^0}$ and which are unassociated to tracks.

There are three important backgrounds. The first is \klpmz\ where both
\pipm\ either decay upstream of the calorimeter (decay in flight) or
punch through to MU3.  The second is \klmthree\ with one decay in
flight or punch-through and accidentally coincident calorimeter
activity that appears as a $\pi^0$.  The third and largest background
is the radiative muonic Dalitz decay \klmumugg\ when
$m_{\gamma\gamma} = m_{\pi^0}$.  Because of the low
expected \cite{ref:fourbody} branching ratio,
{\mbox{$K_L \rightarrow \pi^0 \pi^\pm \mu^\mp \nu$}} is not a
large background.

Two triggers were used for this analysis.  To determine the number of
\kl\ decays in the data, we identified \klpmz\ decays in a minimum
bias trigger.  This trigger required hits in the trigger hodoscopes
and the drift chambers which were consistent with two coincident
charged particles passing through the detector.  Events with a
reconstructed vertex from oppositely charged tracks were recorded with
a prescale factor of 500:1.  For the signal trigger further
requirements were made.  Two or more hits in MU3 were required,
and activity in the photon veto counters rejected events.  The trigger
system counted the number of calorimeter clusters over $\sim1$\gev\ in
a narrow (20nsec) time gate; for the signal mode, at least one such
cluster was required.  We also required that the calorimeter energy
reconstructed online and associated with each track be less than
5\gev.  The signal trigger was not prescaled.

Muons passing through the calorimeter typically deposit $\sim$400\mev,
and so in searching offline for the signal we required that the
clusters associated to the tracks had less than 1\gev\ of energy.
Track momenta were required to be greater than 10 GeV/c to ensure that
they penetrated to MU3 and less than 100 GeV/c to ensure that the
momentum was well measured.  The two-muon system was required to have
a mass less than 350\mevcc\ to reduce backgrounds from \klmthree.  We
required two non-adjacent hits in both views of MU3.  We did not
compare the MU3 hit positions to the extrapolation of the tracks to
MU3, because the major backgrounds passed this requirement as well as
the signal did.

To suppress $\pi^\pm$\ decays in flight, we required that the
reconstructed vertex occurred in the beam volume of the decay region and
had a $\chi^2$\ of 10 for 1 d.o.f. or less, and that the track segments
upstream and downstream of the spectrometer magnet passed within
1\mm\, (94\% signal acceptance) of each other at the bend plane of the
magnet.  The mass of the two unassociated clusters under the hypothesis
that they were produced by photons from the decay vertex was required
to be between 135$\pm$6\mevcc\,($\pm2.5\sigma$).  These
clusters both had to have been found by the trigger cluster counter.

A number of kinematic criteria were studied to suppress background from
\klmumugg\ decays.  The kinematics of this decay are very different
from the analogous \kleegg\ background to \klpiee, and the branching
ratio is lower.  Consequently the methods \cite{ref:Herb,ref:piee}
which are effective in the $e^\pm$\ case are less effective in the
$\mu^\pm$\ case and were not applied.

It is possible to suppress backgrounds from \klpmz, and \klmthree\ by
using
\begin{equation}
{\mathrm R}^{\mu\mu}_{\parallel} \equiv \frac
{(m_K^2 -m_{\mu\mu}^2 -m_{\pi^0}^2)^2 -4 m_{\mu\mu}^2 m_{\pi^0}^2 
                                  -4 m_K^2 p_{\perp \mu\mu}^2}
{p_{\perp \mu\mu}^2 +m_{\mu\mu}^2}
\label{eq:mm0kin}
\end{equation}
where $m_K$\ is the kaon mass, $m_{\mu\mu}$\ the mass of the two-muon
system, $m_{\pi^0}$\ the $\pi^0$\ mass, and $p_{\perp \mu\mu}$\ is the
two-muon systems's momentum perpendicular to the kaon momentum.  This
quantity is proportional to the \piz\ momentum squared in the \kl\
flight direction in the frame colinear with the \kl\ but where the
$\mu^+\mu^-$\ pair has no longitudinal momentum.  We required
${\mathrm R}^{\mu\mu}_{\parallel}$\ to lie between -0.01 and
0.10\,$GeV^2/c^4$.  This cut keeps 89.2\% of the signal and rejects
73\% of \klmthree\ decays with coincident photons and 95\% of the
\klpmz\ decays.

To ensure that we observed all the products of a \kl\ decay, we
required that the total squared momentum transverse to the \kl\ flight 
direction (\itaptsq) be less than 100\,$MeV^2/c^2$, and that the
reconstructed mass ($m$) of the \kl\ be between 492 and 504\mevcc.  With
these requirements, which were selected by examining Monte Carlo
simulation results and data outside the signal region before examining
the data for \klpimumu\ candidates, the overall acceptance for the
signal was 5.0\%.  The simulation distributed the products of the
decay uniformly in phase space.

\begin{figure}
\caption{Reconstructed \itaptsq\ \vs\ $m$\ after all other selection 
criteria, for the data.  The box indicates the signal region; events
in the lower left are predominantly \klpmz\ with $\pi^\pm$\, decays
in flight.}
\label{fig:result}
\epsfig{file=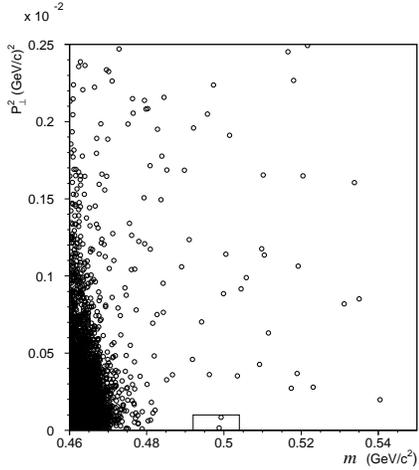,height=2.5in}
\end{figure}

\begin{figure}
\caption{Reconstructed $m$\ after all other selection criteria, for
the data (dots), and the total background estimate (line).  The arrows
indicate the range of values of $m$\ accepted.  The small bump in the
background at the \kl\ mass is from \klmumugg.}
\label{fig:bkg}
\epsfig{file=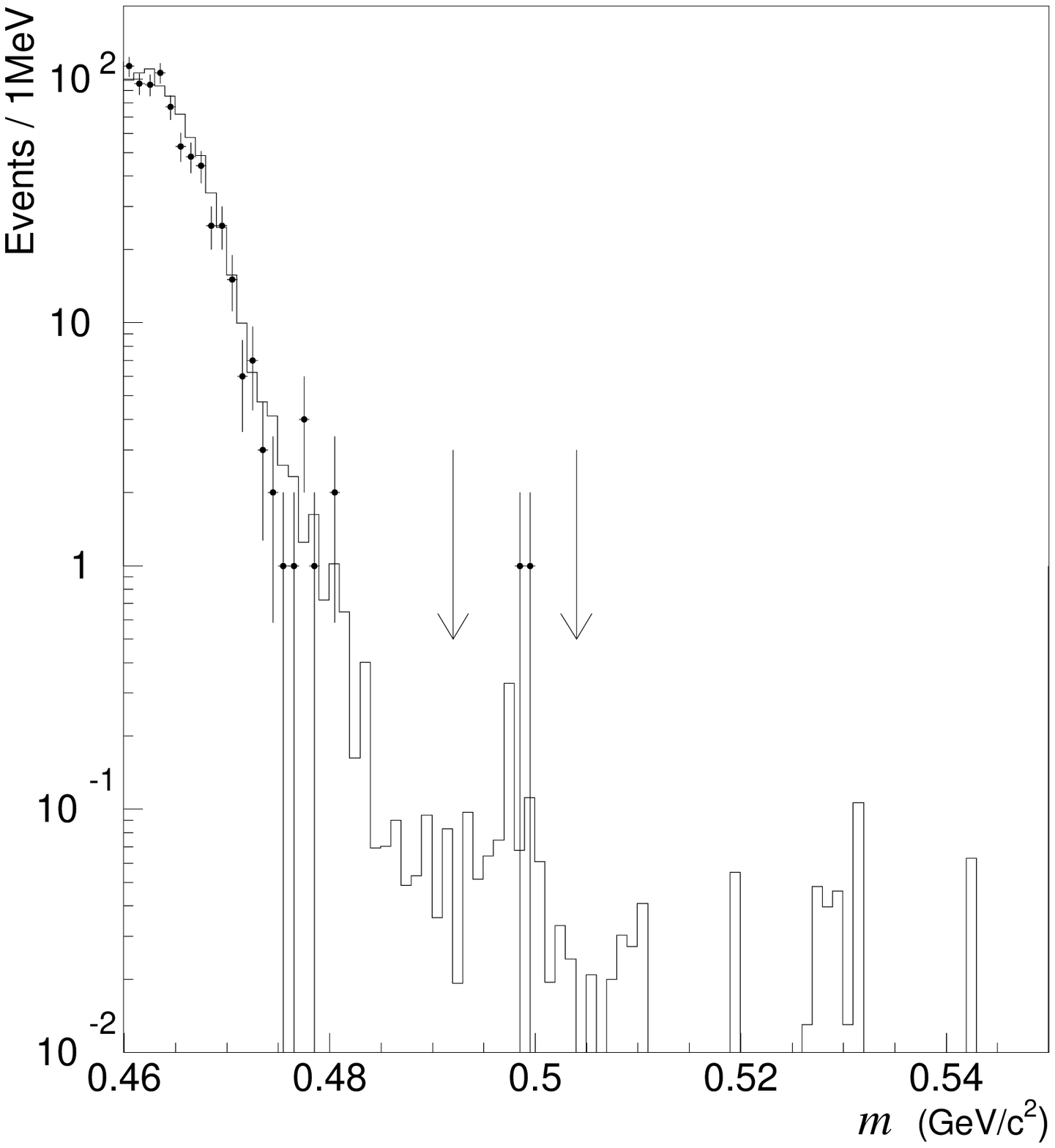,height=3.0in}
\end{figure}

Figure \ref{fig:result} shows the \itapt$^2$\ \vs\ $m$\ distributions
for the data, and Fig. \ref{fig:bkg} shows the mass distribution after
the \itaptsq\ requirement for the data and the backgrounds as estimated
from the simulation.  The correspondence between the data and the
simulation is good, and is also good in the distributions of
$m_{\mu\mu}$, \itaptsq, track momentum, vertex position and (for
\klpmz) of $m_{\gamma\gamma}$.  The background levels in the signal
region are given in Table \ref{tbl:bkg}; they are calculated from the
simulations, published \cite{ref:PDG} branching ratios, and the number
of \kl\ decays in the data sample.  Although we have observed
\klmumugg\ \cite{ref:mmgg} in this data, this background was more
precisely estimated from QED and the measured \BR{\klmumug} ; a value of
\BR{\klmumugg} $\;= (9.1 \pm 0.8) \times 10^{-9}$\ with an
$m_{\gamma\gamma} > 1$\mevcc\ cutoff was used.  All Monte Carlo samples
were over 10 times the data sample.  The background from
\klpm\ + $2\gamma$ (Acc) and
\mbox{$K_L \rightarrow \pi^+ \pi^- \gamma + \gamma$ (Acc)} was
negligible.

To normalize any possible signal's branching ratio, we identified
\klpmz\ decays in as similar a manner to the \klpimumu\ identification
as possible.  Apart from the trigger differences, the calorimeter
energy requirement for clusters associated to tracks was changed to
less than 0.9 times the momentum measured with the spectrometer; the
MU3 and ${\mathrm R}^{\mu\mu}_{\parallel}$\ requirements were removed,
and $m_{\pi^\pm}$\ rather than $m_{\mu^\pm}$\ was used in calculating
kinematic quantities.  The acceptance for the normalization mode was
8.1\%.  There were
$(268 \pm0.4_{\mathrm STAT} \pm0.4_{\mathrm MC} \pm4.3_{\mathrm BR})
\times10^{9}$\, \kl\ decays between 90 and 160\,m from the target with
\kl\ momentum between 20 and 220\gevc.

In calculating the number of \kl\ decays in the data and the acceptance
for \klpimumu\ relative to the acceptance for \klpmz, we allowed for the
uncertainties summarized in Table \ref{tbl:systs}.  We calculated the
\kl\ flux using \klmthree\ rather than \klpmz\ decays and attributed
the difference of 4.20\% to the quality of our simulation of muons in
the detector.  We varied the scale and resolution of the calorimeter
and spectrometer in the simulation to conservatively cover the range of
variations seen in the data.  Apart from uncertainties in published
branching ratios, other sources of uncertainty were small.

From Figs. \ref{fig:result} and \ref{fig:bkg}, two events exist in the
signal region for the data.  Sidebands in both $m$\ and \itapt$^2$\ show
correspondence between data and background predictions as given in 
Table \ref{tbl:sides}.  With the above acceptance and \kl\ flux, and
allowing for a background level of $0.87 \pm 0.15$\ events, we set
\cite{ref:FC} an upper limit \BR{\klpimumu} $\,<3.8 \times 10^{-10}$\
at the 90\% confidence level.

This limit is approximately one order of magnitude more stringent than
the previous limit.  If we assume that the only contribution to the
branching ratio is from direct $CP$\, violation, we may conclude that
$|\eta| < 7$\, at the 90\% confidence level.  In comparison to the
search for \klpiee, \klpimumu\ searches will have better single event
sensitivity for any given sample of \kl\ decays because the level of
irreducible \klelelgg\ background is less.  While not yet as sensitive
as B decays, where a recent indirect global analysis \cite{ref:CKMlims}
finds $\eta$\ to be below 1, it is valuable to test if the same
parameterization is valid for both B and K decays.
%
%
%

We gratefully acknowledge the support and effort of the Fermilab
staff and the technical staffs of the participating institutions for
their vital contributions.  This work was supported in part by the U.S.
Department of Energy, The National Science Foundation and The Ministry
of Education and Science of Japan.  In addition, A.R.B., E.B. and
S.V.S. acknowledge support from the NYI program of the NSF; A.R.B. and
E.B. from the Alfred P. Sloan Foundation; E.B. from the OJI program of
the DOE; K.H., K.S., T.N. and M.S. from the Japan Society for the
Promotion of Science.

\narrowtext
\begin{table}
\caption{Summary of expected backgrounds in the signal region.
``Acc.'' refers to particles from other sources which were accidentally
time-coincident with the interesting decay; ``D'' and ``P'' refer
respectively to decay in flight or punch-through.  Limits are
90\% C.L.; uncertainties are due to uncertainties in published
branching ratios, simulation statistics, and the statistics of the
normalization mode.}
\label{tbl:bkg}
\begin{tabular}{lc}
\multicolumn{1}{c}{Decay mode}       & Expected number of events \\
\tableline
\klmumugg\                           & $0.373 \pm 0.032$         \\
\klmumug\ + $\gamma$(Acc)            & $<0.029$                  \\
\klpmz\ (DD)                         & $0.252 \pm 0.095$         \\
\klpmz\ (DP)                         & $0.007 \pm 0.007$         \\
\klpmz\ (PP)                         & $0.007 \pm 0.007$         \\
\klmthree\ + $2\gamma$(Acc)\  (D)    & $0.161 \pm 0.093$         \\
\klmthree\ + $2\gamma$(Acc)\  (P)    & $0.063 \pm 0.037$         \\
{\mbox{$K_L \rightarrow \pi^0 \pi^\pm \mu^\mp \nu$}} (D)
                                     & $0.009 \pm 0.009$         \\
{\mbox{$K_L \rightarrow \pi^0 \pi^\pm \mu^\mp \nu$}} (P)
                                     & $<0.009$                  \\
\tableline
\multicolumn{1}{c}{Total}            & $0.87 \pm 0.15$           \\
\end{tabular}
\end{table}

\narrowtext
\begin{table}
\caption{Systematic and statistical sources of uncertainty.  Sources marked
with (*) contribute to uncertainty in both the \kl\ flux and the 
acceptance for \klpimumu\ relative to the acceptance for \klpmz; other
sources contribute only to the acceptance ratio.}
\label{tbl:systs}
\begin{tabular}{lc}
\multicolumn{1}{c}{Source}           & Relative Uncertainty \\
\tableline
\BR{\klpmz}                          & 1.59\% (*)           \\
Data statistics for \klpmz\          & 0.16\% (*)           \\
Simulation statistics for \klpmz\    & 0.14\% (*)           \\
Simulation statistics for \klpimumu\ & 0.16\%               \\
Calorimeter scale and resolution     & 3.33\%               \\
Spectrometer scale and resolution    & 1.12\%               \\
Muon identification                  & 4.20\%               \\
Signal trigger requirements          & 0.80\%               \\
Vertex quality requirement           & 0.22\%               \\
Spectrometer wire inefficiency       & 0.15\%               \\
\tableline
\multicolumn{1}{c}{Total}            & 5.77\%               \\
\end{tabular}
\end{table}

\narrowtext
\begin{table}
\caption{Number of observed and predicted events in regions near the
signal region.}
\label{tbl:sides}
\begin{tabular}{lcc}
\multicolumn{1}{c}{Region}                           & Prediction & Observed \\
\tableline
480\mevcc$<m<$492\mevcc; \itaptsq $<$0.0001(\gevc)$^2$ & 2.75$\pm$0.73 &  2 \\
504\mevcc$<m<$515\mevcc; \itaptsq $<$0.0001(\gevc)$^2$ & 0.14$\pm$0.08 &  0 \\
492\mevcc$<m<$504\mevcc; 0.0001(\gevc)$^2<$\itaptsq $<$0.0005(\gevc)$^2$ 
                                                      &  2.65$\pm$1.53  &  2 \\
492\mevcc$<m<$504\mevcc; 0.0005(\gevc)$^2<$\itaptsq $<$0.0010(\gevc)$^2$ 
                                                      &  2.80$\pm$2.59  &  2 \\
\end{tabular}
\end{table}
%
%

\end{document}